# Undamped Soliton-like Domain Wall Motion in Sliding Ferroelectrics

Yubai Shi,[1, 2, *] Yuxiang Gao,[3, 4, *] Hua Wang,[5] Bingwen Zhang,[6] Ri He,[1, 2, †] and Zhicheng Zhong[3, 4, ‡]

[1]*CAS Key Laboratory of Magnetic Materials and Devices, Zhejiang Province Key Laboratory of Magnetic Materials and Application Technology, Ningbo Institute of Materials Technology and Engineering, Chinese Academy of Sciences, Ningbo 315201, China*

[2]*College of Materials Science and Opto-Electronic Technology, University of Chinese Academy of Sciences, Beijing 100049, China*

[3]*School of Artificial Intelligence and Data Science, University of Science and Technology of China, Hefei 230026, China*

[4]*Suzhou Institute for Advanced Research, University of Science and Technology of China, Suzhou 215123, China*

[5]*Center for Quantum Matter, School of Physics, Zhejiang University, Hangzhou 310058, China*

[6]*Fujian Key Laboratory of Functional Marine Sensing Materials, Center for Advanced Marine Materials and Smart Sensors, College of Material and Chemical Engineering, Minjiang University, Fuzhou 350108, China*

Sliding ferroelectricity in bilayer van der Waals materials exhibits ultrafast switching speed and fatigue resistance during the polarization switching, offering an avenue for the design of memories and neuromorphic devices. The unique polarization switching behavior originates from the distinct characteristics of domain wall (DW), which possesses broader width and faster motion compared to conventional ferroelectrics. Herein, using machine-learning-assisted molecular dynamics simulations and field theory analysis, we predict an undamped soliton-like DW motion in sliding ferroelectrics. It is found that the DW in sliding ferroelectric bilayer 3R-$MoS_2$ exhibits uniformly accelerated motion under an external field, with its velocity ultimately reaches the relativistic-like limit due to continuous acceleration. Remarkably, the DW velocity remains constant even after the external field removal, completely deviating from the velocity breakdown observed in conventional ferroelectrics. This work provides opportunities for applications of sliding ferroelectrics in memory devices based on DW engineering.

---

[*]These authors contributed equally to this work.
[†]heri@nimte.ac.cn
[‡]zczhong@ustc.edu.cn

Sliding ferroelectricity, recently discovered in 2D van der Waals (vdW) materials such as bilayer hexagonal boron nitride (h-BN) and molybdenum disulfide (3R-$MoS_2$) [1-3], offers significant potential for nonvolatile memory applications. This potential stems from its robustness against depolarization fields at ultrathin thicknesses and excellent fatigue resistance during polarization switching cycles [4-8]. Different from conventional ferroelectrics where polarization stems from ionic displacements, the polarization in sliding ferroelectrics originates from distinct bilayer stacking configurations in nonpolar parent materials [1,2]. With polarization switching facilitated by in-plane interlayer sliding [9-11], sliding ferroelectrics exhibit significantly ultrafast switching dynamics compared to conventional ferroelectrics under both electric fields and laser pulses [5,6,12-14], thereby offering prospects for enhanced read/write speeds nonvolatile memory devices.

Polarization switching is typically achieved through the domain wall (DW) motion [15,16]. DW in ferroelectrics and ferromagnets is boundary between regions with different polarization and magnetic orientations, respectively. It is a soliton with localized energy, capable of being driven by an external field, whose motion can be analogized to that of a classical particle and exhibits inertia in principle [17-20]. However, in conventional systems, DW motion is accompanied by damping, leading to a uniform terminal velocity under an external field and violent dissipation of motion upon field removal [21-26]. Consequently, continuous and large external field is required for microelectronic devices based on DW engineering, such as racetrack memories, which in turn causing significant power consumption issues [27-29]. In sliding ferroelectrics, the DW exhibits exceptionally high mobility with velocities exceeding those in conventional ferroelectrics and ferromagnets by two orders of magnitude, serving as a promising candidate to overcome dissipation issues and enhance device performance [5,6,30]. Therefore, a comprehensive understanding of the damping behavior governing DW dynamics in sliding ferroelectrics is crucial and requires further investigation.

In this Letter, using large scale atomistic simulations based on the machine learning interatomic potential within density functional theory (DFT) accuracy [31], we study the dynamics of DWs in a typical sliding ferroelectric bilayer 3R-MoS$_2$. Our results reveal that the DWs in this system exhibit undamped motion, with their velocity remaining constant even after the removal of external field. Employing field theory, we analytically derive equation for DW motion, which can be interpreted as Newton's second law. We also demonstrate that under continuous electric field driving, DW velocity can reaches the relativistic-like limit, which is the speed of in-plane transverse acoustic phonon mode in MoS$_2$. Additionally, we propose an experimental framework to validate this unique soliton-like DW phenomenon.

We start with the stacking configurations of bilayer 3R-MoS$_2$. There are stable XM and MX stacking modes, as shown in Fig.1(a). The S (Mo) atoms in the top layer of XM (MX) state are aligned with the Mo (S) atoms in the bottom layer. The XM state can transfer to MX state by the interlayer relative sliding, overcoming the energy barrier. In the sliding process, interlayer charge transfer occurs, which leads to the opposite out of plane spontaneous polarization of XM and MX states. [see Fig. 1(b)] [2]. These results are calculated by DFT (see Supplemental Material S1; also see Refs. [15,31-39] therein), consistent with previous studies [5,11]. In conventional ferroelectrics, polarization switching is typically driven by domain wall motion, and the same is true for sliding ferroelectrics. To study the motion of DW in bilayer 3R-MoS$_2$, we construct a freestanding supercell with over 30,000 atoms, containing two DWs. Based on the developed Deep Potential (DP) model of MoS$_2$ [5], we perform atomic relaxation to obtain the optimized DW structure [see Fig. 1(c)]. The pink dots plot the variation of sliding distance ($u_s$) along the $x$-axis. The DW has a characteristic width of approximately 100 Å, significantly larger than that in conventional ferroelectrics (~8 Å) [16]. The polarization orientation gradually rotates in $yz$-plane, forming a Bloch-like DW. The above results are similar to the DW in $h$-BN, indicating the universal property of DW in sliding ferroelectrics [30].

The formation of DW in sliding ferroelectricity is governed by the competition between interlayer stacking energy and elastic energy associated with in-plane lattice distortion. These two energy terms can be written as functions of the sliding distance. The interlayer sliding distance is treated as a (1+1)-dimensional classical scalar field $u_s(x, t)$, and analyzed by field theory to further understand the properties of the DW from the fundamental physical perspective. The soliton solution of sine-Gordon equation is as follows (for detail formula derivation, see Supplemental Material S2):

$$u_s(x,t) = \frac{2u_0}{\pi} \tan^{-1}\left\{\exp \pm \left[\frac{\pi\sqrt{2\Delta/\lambda}\,(x - x_0 - vt)}{\gamma u_0}\right]\right\}, \qquad (1)$$

where $\Delta$ represents the barrier height in Fig. 1(a) and $u_0$ the sliding distance from XM to MX state. $x_0$ is location of the soliton center. And $\gamma = \sqrt{1 - \beta^2}$, $\beta = v/v_c$, $v_c = \sqrt{\lambda/\rho}$, where $\rho$ is the mass density and $\lambda$ the coefficient of energy cost with in-plane lattice distortion. $v$ is velocity of soliton and $v_c$ is actually the speed of in-plane transverse acoustic phonon mode in $MoS_2$, which we will discuss in the next part. First examine the static solution, when the $v$ equals zero (see Eq. S4). Here, soliton (+) and antisoliton (−) of field $u_s$ represent the ferroelectric DWs in bilayer 3R-$MoS_2$. Tthe energy distribution of the DW can be further determined by Eq. S5. Finally, by substituting the values of $\rho$, $\lambda$, $\Delta$, and $u_0$ (The values and calculation procedures are in Supplementary Material S2) into Eq. (1) and Eq. S5, the waveform of the soliton and energy distributions in the domain structure are obtained, as shown by the blue and green lines in Fig. 1(c), in agreement with the results of DP model. The $u_s$ has localized energy distribution at the DW position, DW energy can be obtain by integration (see Eq. S6). We attribute the slightly narrower domain wall width and the slightly lower domain wall energy predicted by DP compared to those calculated by the field theory model based on DFT results to the intrinsic cutoff radius of DP. The cutoff radius leads to the neglect of long-range interactions, causing a partial loss of the energy associated with in-plane lattice distortions, which leads to minor bias with the results of field theory model.

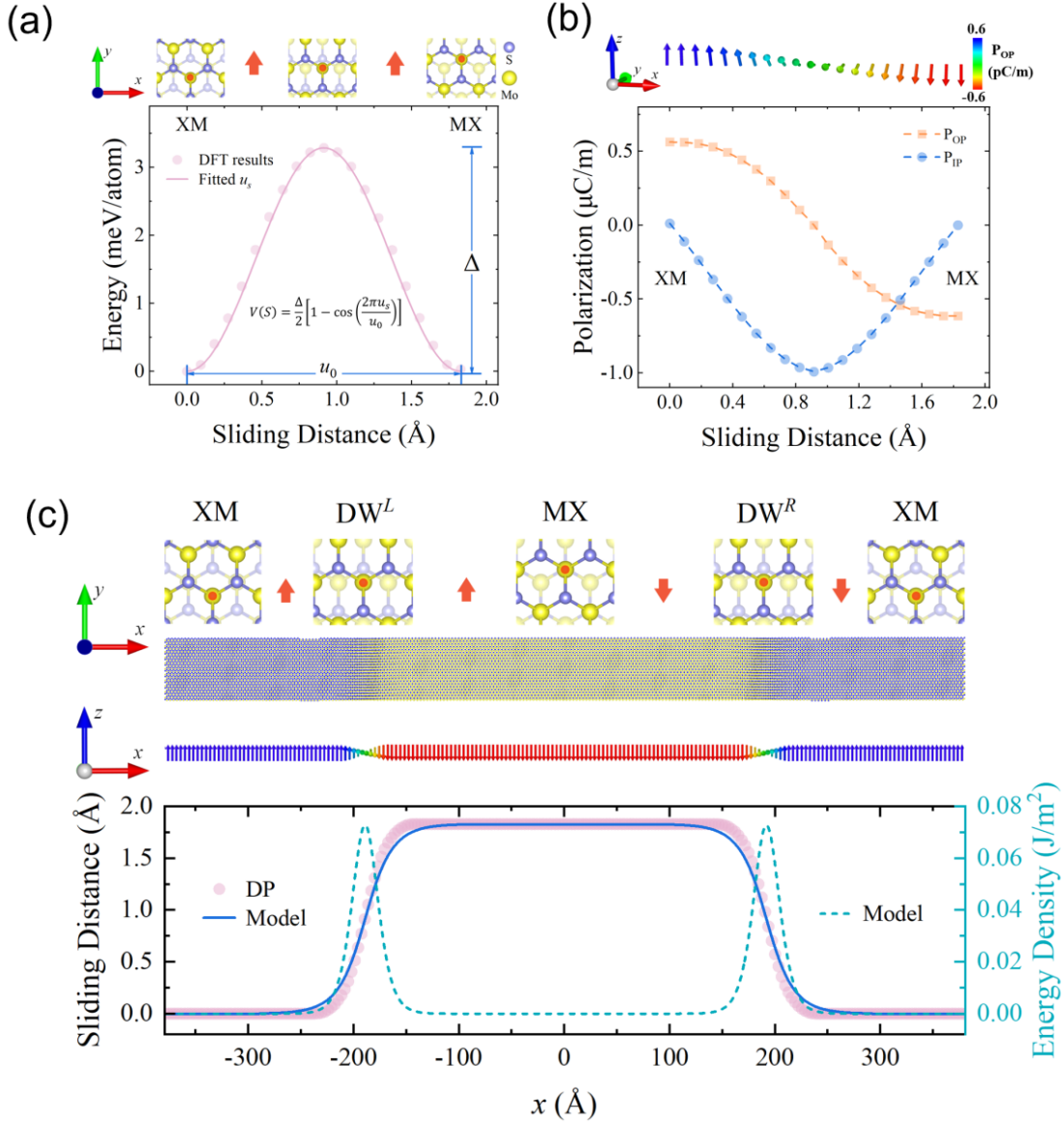

FIG. 1. (a) Stacking mode and sliding barrier of bilayer 3R-MoS$_2$, calculated by density functional theory. The marked Mo atom and red arrows show the interlayer sliding orientation. We use cosine function to fit the potential energy curve. $\Delta$ is the barrier height and $u_0$ is the sliding distance from XM to MX state. (b) The polarization variation in the sliding process, calculated by density functional perturbation theory. (c) The static domain wall structure of bilayer 3R-MoS$_2$. The pink dot is distribution of sliding distance in real space after atomic relaxation, calculated by DP model. The blue line and the green line are the sliding distance and the energy distribution of the domain wall, fitted by field theory model, respectively.

We further focus on the dynamic properties of the DW under an external field. To mitigate the effects of thermal fluctuations, molecular dynamics simulation is performed at 1 K (computational details see Supplemental Material S1). The time-dependent displacement of DW$^L$ and DW$^R$ under different constant electric field are shown in Fig. 2 (a) and (b). $DW^L$ and $DW^R$ exhibit opposite motions: under a $+z$ ($-z$) field, the blue (yellow) domain with $+z$ ($-z$) polarization expands, and the DW moves toward to the yellow (blue) domain. The moving DW corresponds to the time-dependent soliton solution of undamped sine-Gordon equation [see Eq. (1)]. The critical velocity $v_c$ is the speed of in-plane transverse acoustic phonon mode in $MoS_2$, which is related to in-plane lattice stiffness and mass density and is estimated to be on the order of 3000 m/s [40]. It is the ultimate physical speed limit for DW velocity $v$, playing a role similar to the speed of light in special relativity, which means $v$ must be less than $v_c$ in magnitude. The DW velocity will exhibit relativistic-like effects as increasing to $v_c$, similar to previous findings in ferromagnetic DWs [18]. And the energy of the moving DW can be written as: $\mathbb{E}_v = \mathbb{E}/\gamma$. This solution can be regarded as an extended free particle moving at velocity $v$ with energy $\mathbb{E}_v$. $\mathbb{E}_v$ exhibits the right relativistic dependence on the $v$ of the soliton. Therefore, at a low velocity ($v \ll v_c$), the configuration of the energy density only undergoes a central translation without deformation.

Importantly, in the low velocity regime ($v \ll v_c$), the displacement exhibits a clear quadratic relationship dependence on time, while the DW velocity increases with the strength of the strength of electric field. Therefore, we fit the displacement curve with a quadratic function and found that the DW exhibits constant acceleration under a specific electric field, demonstrating uniformly accelerated motion (The fitted figures and data are in Fig. S2-3 and Table S1-2). As shown in Fig. 2(c), the acceleration of the DW exhibits a linear dependence on the field strength. These results indicate that the DW with a low velocity ($v \ll v_c$) behaves as classical particles with their dynamics governed by Newton's second law. The same result can be obtained by deriving sine-Gordon equation in the presence of both external driving forces and damping term, which can be treated by linear perturbation theory (see Supplemental Material S2 and Ref. [20]). The kinetic equation of soliton is as follows:

$$\frac{\partial^2 \phi_b}{\partial t^2} + \frac{\pi\gamma\Gamma}{u_0}\sqrt{\frac{2\Delta}{\rho}}\frac{\partial \phi_b}{\partial t} = \frac{16\beta\gamma\Gamma\Delta}{\rho u_0^2} + \frac{4\pi^2\sigma}{\rho u_0}E \qquad (2)$$

where $\phi_b$ is amplitude of the translational mode. The position of DW center is proportional to $\phi_b$. When the damping coefficient $\Gamma$ approaches zero, the formula reduced to Eq. S17. $4\pi^2\sigma/\rho u_0$ plays the role of the "mass-to-charge ratio" of the DW, which represents the response intensity of DW motion to the electric field. $\sigma$ should be treated as a tensor. The direction of DW motion occurs perpendicular to the external field, resulting in nonzero off-diagonal element of $\sigma$. Finally, we obtain an equation in the form of undamped Newton's second law (Eq. S17), that the second time derivative of the DW position is constant, in excellent agreement with the DP results in Fig. 2(d). Additionally, molecular dynamic simulations with the larger structure (see Supplemental Material S1) and electric field [three orders of magnitude larger than the values in Fig. 2(d)] is performed to validate the relativistic-like effect as DW velocity increases to $v_c$. As shown in Fig. 2(e), under an electric field of 0.56V/nm, the DW velocity increases linearly with time initially, exhibiting uniformly accelerated motion. As the velocity increases and approaches $v_c$, the acceleration gradually decreases. Eventually, the velocity ceases to increase at around 4000 m/s, which is close to the value estimated from our field theory analysis above, indicating the onset of relativistic-like effects.

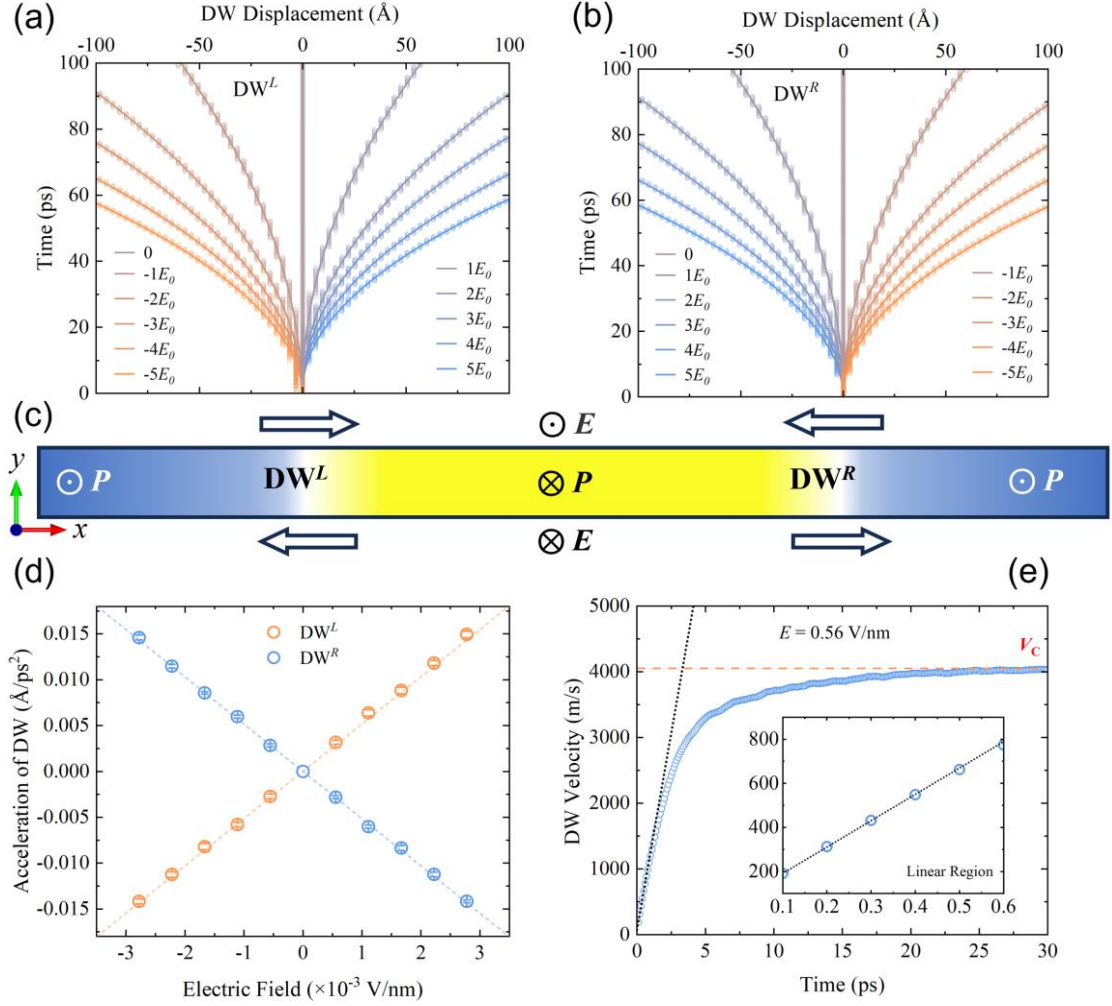

FIG. 2. (a) Evolution of displacement of left DW (DW$^L$) and (b) right DW (DW$^R$) under different constant external electric field, where $E_0$ equals 5.6 ×10$^{-4}$ V/nm. DW$^L$ and DW$^R$ exhibit opposite motions. (c) A real-space schematic of DW motion. (d) The acceleration of DW$^L$ and DW$^R$ under different constant external electric field. The acceleration exhibits a linear relationship with the field strength. (e) The time-dependent evolution of DW velocity under a large electric field [three orders of magnitude larger than the values in (d)]. DW exhibits relativistic-like effects as its velocity approaches $v_c$.

Such undamped motion mode of DW in sliding ferroelectrics differs significantly from that of the one in conventional ferroelectrics, where the DW usually maintain a constant velocity under a specific external electric field, following the creep equation [41-43]. This is due to the presence of the damping term ($\frac{\pi\gamma\Gamma}{u_0}\sqrt{\frac{2\Delta}{\rho}}\frac{\partial\phi_b}{\partial t}$) with a first-order time derivative in the equation of motion [see Eq. (2)]. It balances with the external field, allowing the DW to reach a terminal velocity. Moreover, when the external field is removed, the velocity will decrease to zero immediately [21,23]. Therefore, we further investigate the dynamic behavior of DW in sliding ferroelectrics after the external field is removed. As shown in Fig. 3(a), and (d), the DW in $MoS_2$ continue to move at a constant velocity even after the external field turned off. Temperature is increased to 300 K to assess the impact of thermal fluctuations. The results show that the DW remains in motion after the electric field is removed [see Fig. 3(b)], exhibiting the same behavior at 1 K. This demonstrates that thermal fluctuations only induce low dissipation in DW motion. The DW motion remains consistent with our results in Fig. 2 and Eq. (2), exhibiting undamped motion like a free Newtonian particle. The behavior of 180° DW in conventional ferroelectric $PbTiO_3$ under an electric field at 300 K is also examined for comparison [see Fig. S1, Fig. 3(b) and (c)]. The DW velocity remains constant under the electric field, but it decreases to zero once the field is removed, in good agreement with previous studies [21].

The dissipation of DW motion in different materials originates from various damping mechanisms. For example, the Gilbert damping in ferromagnets is caused by spin-orbit and spin-lattice coupling [44]. In ferroelectrics, several studies have attributed the damping of DW motion to its coupling with acoustic phonons, where energy is dissipated into the lattice thermal reservoir through phonon scattering [45,46]. For conventional ferroelectrics, their DW motion is coupled with the vibration modes, which are transmitted by ionic bond. While in sliding ferroelectrics, the DW motion interacts with the interlayer shear mode, as the polarization switching is facilitated by interlayer sliding. The shear mode transmits through the weak interlayer van der Waals forces, with a frequency much lower than other vibration modes [47]. Therefore, it is less affected by phonon scattering, which significantly reduces energy dissipation caused by lattice vibrations during domain wall motion, therefore leads to the almost zero damping.

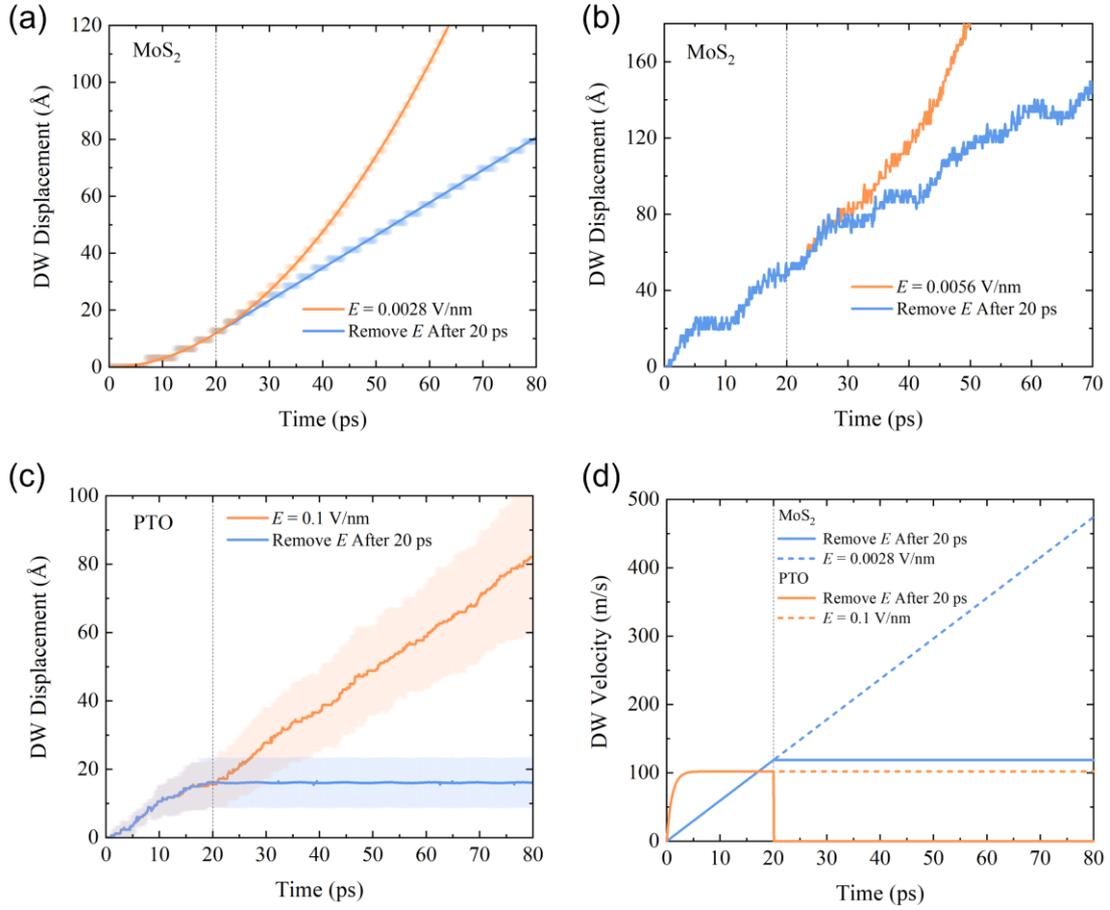

FIG. 3. DW displacement evolution of bilayer 3R-MoS$_2$ at (a) 1 K and (b) 300 K, (c) PbTiO$_3$ at 300 K. The electric field is turned off after 20 ps. (d) DW velocity evolution of bilayer 3R-MoS$_2$ and PbTiO$_3$. After the electric field is turned off at 20 ps, DW in MoS$_2$ can still maintain an inertial motion with constant velocity, while the one in PTO becomes stationary.

The special DW motion in sliding ferroelectrics can be experimentally detected through time-delayed signals using non-local devices. By applying an external field pulse or even a laser pulse at a specific location and time to drive the DW motion, and then observing the polarization switching at a different location and time, this effect can be monitored. As shown in Fig. 4, we propose an experimental scheme to verify the undamped DW motion in bilayer 3R-MoS$_2$. One end of the bilayer 3R-MoS$_2$ film is connected to a current source via insulating layer to suppress leakage currents, while the other end is directly connected to electrodes for voltage detection. When a voltage pulse is applied at $t_1$, DW motion can be triggered. In the ideal case, the DW maintains uniform motion. Even in the presence of defects, such as sulfur vacancies, it can also continue to move if it has already achieved a sufficiently high velocity [5]. After DW reaches the opposite end of the film, polarization switching can induce a signal of change in voltage at $t_2$. The detected signal is delayed relative to the applied voltage pulse with time interval $\Delta t = t_2 - t_1$, providing evidence of the undamped DW motion the undamped motion of DW after the removal of an external electric field. The pulse width needs to be on the picosecond scale, as the film length is on the micrometer scale and the DW velocity is approximately ~km/s. Consequently, $\Delta t$ is on the nanosecond scale, which can be effectively distinguished in experiment.

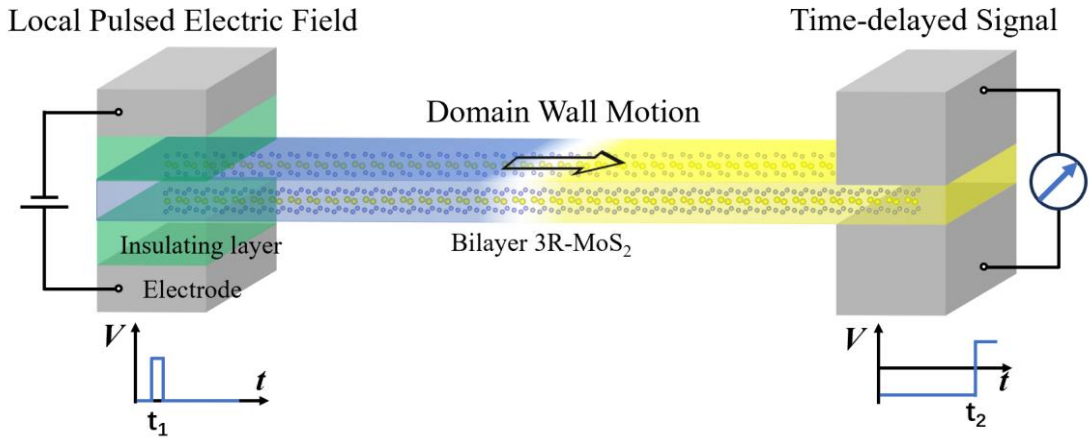

FIG. 4. Device geometry for verifying special DW motion in sliding ferroelectrics. The electric field pulse is applied on the left side of the $MoS_2$ film and the polarization switching signal is detected on the right side of the film.

This unique behavior of DWs in sliding ferroelectrics open new avenues for the design of neuromorphic devices or non-volatile memories like racetrack memories, which traditionally rely on magnetic DWs [28]. There are several challenges, such as DWs being easily pinned by defects, leading to read-write fatigue. Furthermore, the magnetic DW velocity is at ~100m/s range which limits the read/write speed. More critically, the uninterrupted and large spin current is required during read-write operation, resulting in excessive power consumption and associated heating issues [29]. In contrast, designing racetrack memory based on sliding ferroelectrics holds promise for overcoming these challenges. Sliding ferroelectrics are compatible with current complementary metal-oxide-semiconductor technologies and exhibit fatigue resistance during polarization switching cycles [5]. The DW in these materials exhibits ultrafast motion (~km/s), which can significantly enhance the read-write speed. Due to its undamped behavior, only electric field pulses are required for operation, eliminating the need for continuous current. This approach not only reduces power consumption but also mitigates heating issues, making it an attractive alternative for future memory technologies.

In summary, by combining DFT calculations, machine-learning interatomic potential based molecular dynamics simulations and field theory analysis, we studied the static structure and dynamic properties of DW in bilayer 3R-$MoS_2$. We predict an undamped motion of DW in sliding ferroelectrics, a phenomenon contrasting with the damped motion in conventional ferroelectrics. The DW in sliding ferroelectrics exhibits uniform acceleration under an external electric field, just like the free soliton which obeys Newton's second law. Moreover, the DW motion exhibits relativistic-like effect as its velocity approaches $v_c$, which is the speed of in-plane transverse acoustic phonon mode in $MoS_2$. We highlight the steady inertial motion of DW in sliding ferroelectrics after the external field removal, which has not been observed in conventional ferroelectrics to our best knowledge. This work thereby benefits the application of sliding ferroelectrics in microelectronic devices, e.g., the implementation racetrack memories and neuromorphic devices.

*Acknowledgments*—This work was supported by the National Key R&D Program of China (Grants No. 2022YFA1403000 and No. 2021YFA0718900), the National Nature Science Foundation of China (Grants No. 92477114, NO.12204496, NO. 12374096, No. 12304049, and No. 12474240), the Zhejiang Provincial Natural Science Foundation (Grants No. Q23A040003 and No. LDT23F04014F01), and Ningbo Nature Science Foundation (No.2023J360).